\documentstyle[prb,aps]{revtex}
\tightenlines
\begin{document}
\draft
\input{psfig}
\title{
Pressure-Induced Interlinking of Carbon Nanotubes
}
\author{T. Yildirim$^{(1)}$, O. G\"{u}lseren$^{(1,2)}$,
\c{C}. K{\i}l{\i}\c{c}$^{(3)}$, and S. Ciraci$^{(3)}$}
\address{$^{(1)}$ NIST  Center for Neutron Research, National Institute of Standards
and Technology, Gaithersburg, MD 20899 }
\address{$^{(2)}$ Department of Materials Science, University of Pennsylvania, Philadelphia, PA 19104}
\address{$^{(3)}$ Physics Department, Bilkent University, Ankara, Turkey}
\date{\today}
\preprint{To be appear in PRB}
\maketitle

\begin{abstract}

We predict new forms of carbon consisting of one and  two 
dimensional networks of interlinked  single wall carbon nanotubes,
some of which are energetically more stable than  
{\it van der Waals} packing of the
nanotubes on a hexagonal lattice. 
These interlinked nanotubes are further transformed
with higher applied external pressures
to more dense and complicated stable structures, in which 
curvature-induced carbon sp$^{3}$ re-hybridizations are formed.
We also discuss the energetics of the bond formation
between nanotubes and the electronic properties of 
these predicted novel structures.

\end{abstract}




Carbon nanotubes,
originally discovered  as by-products of 
fullerene synthesis\cite{iijima1,iijima2},  are now 
considered to be the  building blocks of  future
 nanoscale   electronic and mechanical devices. It is therefore
desirable to have a good understanding of their electronic
and mechanical properties and the interrelations between them.
In particular, single wall carbon nanotubes (SWNT) provide
a system where the electronic properties can be controlled
by the structure  of the nanotubes and by various deformations
of their geometries\cite{mint,ellipse,oguz}. 
The physical  properties can also be  altered by  intertube interactions
between  nanotubes packed in  hexagonal lattices,
as  so called ''nanoropes". 

The intertube interactions in nanoropes can be probed  
by applying  external pressure to vary the intertube 
distance\cite{wood,venkateswaran,chesnokov}. 
For fullerenes, such high pressure studies
have yielded many interesting results including  new compounds 
such as the pressure-induced polymeric phases of C$_{60}$\cite{rao}. 
It is, therefore,
of interest to inquire if similar covalent-bonding
can occur between the nanotubes in a rope.
This could have  important consequences for  
nanoscale device applications and composite materials
which require strong mechanical properties since  
nanoropes consisting of inter-linked SWNT will be significantly
stronger than nanoropes composed of van der Waals packed 
nanotubes\cite{salvatat}. 

A recent Raman study on SWNT ropes carried out up to
25.9 GPa\cite{venkateswaran} showed  that the mode intensities and 
energies are not completely reversible upon pressure cycling, 
suggesting  irreversible pressure-induced changes in the structure. 
In another  high pressure study 
Chesnokov {\it et al.}\cite{chesnokov}
observed a very large volume reduction and high 
compressibility, signaling the presence of a 
microscopic volume-reducing deformation other than
van der Waals compression. Some of these
pressure-induced effects are tentatively attributed
to possible crushing or flattening  the nanotube cross
section from circular to elliptical or hexagonal\cite{chesnokov}.
Motivated by these reports,
we investigated possible new pressure-induced ground state structures for
(n,0) nanotube ropes\cite{nmtubes} from  first-principles total energy calculations using the
pseudopotential method within the 
generalized gradient approximation (GGA)\cite{gga}.
We  observed an elliptical distortion of the nanotubes under pressure
and subsequent  curvature-induced carbon re-hybridization, 
giving rise to
one or two dimensional interlinked  networks of nanotubes.
This is somewhat reminiscent to the pressure-induced polymerizations
of C$_{60}$\cite{rao}.

The first principles total energy and electronic structure
calculations were carried out using the pseudopotential 
plane wave code  CASTEP\cite{castep}. 
We used  plane  waves with an energy cutoff of 500 eV. With this cutoff 
and using ultra soft pseudopotentials for carbon\cite{usp}, 
the total energy  converges within 0.5 meV/atom.
For the Brillouin zone integration, we used between  3x3x6 to 5x5x7  k-points
according to the Monkhorst-Pack special {\bf k}--point scheme\cite{kpts}.
This method has already been applied to many carbon 
systems, including fullerenes and cubane with remarkable
success\cite{cubane}. 

In this Letter, we present calculations on nanoropes 
consisting of (5,0),(6,0), (7,0),
(9,0), and (6,6) nanotubes. For simplicity, we model
the nanoropes as a hexagonal   lattice of nanotubes. 
We further assume that we have only one nanotube per 
unit cell.   
(6,0) nanotubes are perfectly compatible with the hexagonal
lattice (i.e. $\gamma = 120^{\rm o}$). However 
in the optimized structure, hexagonal carbon rings
of one nanotube face   hexagonal rings from the 
neighboring tubes (Fig.~1(a)).
From  studies of fullerenes, we know that the 
energy  can be reduced  by rotating  every other
tube such that their C--C bonds face the center of the
hexagonal faces of the neighboring nanotubes. This would 
double the unit cell for (6,0) nanotubes. We will   not
pursue this here because we are mainly interested in  
the potential for covalent bonding between nanotubes.

(9,0) nanotubes are also compatible with  hexagonal symmetry.
Furthermore the relative orientation of the nanotubes are 
optimal energetically  (i.e. C--C bonds 
face the center of the hexagonal faces of  adjacent  nanotubes).  
(5,0) and (7,0) nanotubes are not compatible with  hexagonal 
symmetry and therefore $\gamma$ is expected to deviate from 
the  ideal value of $120^{\rm o}$.
We find that when one bond is along the a-axis, the total 
energy is minimized when   the b--axis
is aligned so  that it brings the C-C bonds of one nanotube
to the center of hexagonal faces of adjacent  nanotubes.
For  (5,0) and (7,0) nanotubes this occurs when 
$ \gamma = (360/5)\times 1=72.0^{\rm o}$
and $(360/7) \times 2 = 102.857^{\rm o}$, respectively. 
These values are very close to the values,
$\gamma=72.46^{\rm o}$ and $102.35^{\rm o}$, obtained
from the first principles structural optimization.

The pressure dependence  of these lattices of nanotubes  
was determined by  calculating the total energy  
as a function of nanotube separation (i.e. $a$ and $b$)
while the other parameters, including atom positions, 
$c$,   and $\gamma$ are optimized.
We observe that  (7,0) nanotubes become
elliptically distorted with applied pressure
(i.e.  decreasing nanotube-nanotube distance). 
At  a critical pressure, we observe a structural phase transformation from 
the van der Waals nanotube lattice (as shown in Fig.~1 a-b) 
to a new lattice
in which the nanotubes are interlinked along the [110] direction, 
where the strain of the nanotube is  largest  (Fig.~1(c)).
The covalent bonding between nanotubes is  therefore
the result of curvature-induced re-hybridization of the carbon orbitals.
The same structural transformation was observed for the  
other (n,0) nanoropes and the results  are summarized in Table~1.  
The structure of each of these one-dimensionally interlinked nanoropes 
is orthorhombic (space group Cmcm), with  two nanotubes
per unit cell. The relationship between the conventional 
orthorhombic cell and the primitive one is shown in Fig.~2(a).

Figure~2 also shows the local environment of the carbon 
atoms involved in the
interbonding of the nanotubes for (6,0) and (7,0). In the first
case, the covalent bonding occurs between carbon atoms on two hexagonal
rings 
(Fig.~2(b)). We believe this process is slightly
less favorable than  covalent bonding between
carbons on a hexagonal  ring and those involved in an intratube C--C bond 
which occurs for the (7,0) case (Fig.~2(c)).
In both cases, the bond
distances are comparable to those in diamond, indicating sp$^{3}$
hybridization. The bond angles  vary from 
about 100$^{\rm o}$  to $120^{\rm o}$, indicating some
strain.
From Table~1, we see that the energy of the  interlinked 
phase is actually  lower than the vdW
lattices for (5,0) and (7,0) nanotropes.  For (9,0) nanorope,
the energy difference is relatively small. 
Regardless of the energies relative to that of unlinked  nanotubes,
an interlinked  phase is stable  once it is formed,
because breaking the intertube bonds 
requires jumping over a significant energy barrier.

To quantitatively  study the bonding mechanism,
we  calculated the total energies of the different
phases as a function of the lattice constant (i.e. applied
pressure). The result for (7,0) nanotubes is summarized in Fig.~3.
The energies  of the vdW and the one dimensional interlinked (1DI)
phases  cross each other  at about a = 9.0 \AA $\;$
with an energy barrier of only 46 meV/unitcell (552~K). 
The pressure required to attain  this lattice constant 
is only about 0.3 GPa for the vdW phase, indicating 
that polymerization of vdW (7,0) nanoropes could occur at 
modest pressures and temperatures\cite{barrier}. Once
the interlinked  phase is reached,  the 
energy barrier required to break the bonds and obtain free nanotubes
is about 0.7 eV (25 meV/atom),  which is comparable to 
that of  1D polymerized C$_{60}$ molecules (20 meV/atom)\cite{adams}.

Fig.~3 also shows that another interlinked phase of (7,0) nanotubes
becomes the ground state for lattice parameter 
smaller than   8.0 \AA .
In this  new phase  the nanotubes 
are interlinked along both a-- and b-- axes
 (see Fig.~4(a)).
This two dimensional interlinked  (2DI) structure is 
about four times stiffer (i.e. $d^{2}E/da^{2} = 13. 7 eV/\AA^{2}$) than
the 1D interlinked phase ($d^{2}E/da^{2} = 3.3 eV/\AA^{2}$)
and sixteen times stiffer than the vdW nanoropes ($d^{2}E/da^{2} = 0.8
eV/\AA^{2}$).

We observe that applying even higher pressures
yields more complicated and denser phases for 
many of the  nanoropes  studied here
(see Fig.~4).  For (9,0) nanoropes,
we find that the nanotubes are interlinked along
three directions forming a hexagonal network. The
length of the intertube bond, $d_{C-C} = 1.644 $ \AA, 
 is significantly elongated for an sp$^{3}$
C--C  bond.
The two dimensional interlinked phase of (7,0) nanotubes
is further transformed to a denser structure at 30 GPa (Fig.4(c)). 
By comparison, (6,6) nanotubes do not form an interlinked 
structure up to a pressure of 60 GPa. 
Rather the  nanotubes
are hexagonally distorted such that the local 
structure of the nanotube faces is  similar to that in  graphite
sheets (Fig.4(c)). Furthermore, releasing the  pressure 
yields the original structure, indicating that the  
distortion is purely  elastic. 
Similar calculations are currently underway 
for other (n,m) nanotubes. However, we expect similar
results for other types of tubes.

A detailed discussion of the electronic band structure and 
density of states of the
predicted structures  will be presented elsewhere, 
however all the structures reported here (except one in Fig.~4(c))
are found to be metal. The dispersion of the bands near Fermi level
in a direction perpendicular to the tube axis  is found to be around 0.5 eV, making
the vdW nanoropes metallic even though the individual tubes are insulating
(such as (7,0) nanotubes) or semiconducting with a small band gap.
The structural changes   clearly have strong effects on
the electronic properties and therefore should be detected in the
pressure dependence of various
transport properties of nanoropes. From our
calculations we expect two effects to be observed.
First, on the initial pressure  cycle one 
should observe  an irreversible effect in the resistivity due
to the polymerization of the nanotubes. Second,
the applied pressure will induce a reversible deformation
of the circular cross section, which will change the metallic
behavior of the nanotubes\cite{oguz}. 
For example, the (7x0) nanorope, which  is found to be metallic  
at ambient pressure, transforms to a high density phase as shown
in Fig.~4(c) at 30 GP, which is  found to be a band insulator 
with a band gap of 2 eV.

The new pressure--induced, high density
phases reported here can   also be  important host
lattices for intercalation and sorption. 
For instance, the high density phase of (7,0) nanoropes (Fig.~4(c))
has an interesting structure in which some parts of the lattice
are very strained containing  sp$^{3}$ bonded square carbon
rings, while other parts consist of  graphite-like sp$^{2}$ 
bonded  carbon atoms (28\% of all atoms). 
In this structure, there are still  interstitial 
sites which would  accommodate  other species such as 
alkali metals. 

In summary, we have presented first-principles calculations
of the structures and electronic properties of various nanoropes.
We find that small nanotubes in a rope are  distorted
elliptically with applied pressure 
and then  are covalently bonded to each
other at the positions of highest curvature point 
of nanotubes 
by sp$^{3}$ hybridization of the carbon 
orbitals. For small nanotubes, the resulting one-dimensional
chains of elliptic carbon nanotube structure are  found
to be energetically more stable than the circular 
van der Waals nanoropes. Higher applied  pressures resulted
in more dense and complicated structures.
Thus pressure induced  polymerization of the nanotubes
may provide a way of synthesizing novel carbon base materials 
with  interesting physical properties.
For example interlinking of the nanotubes may improve
the mechanical performance of composites based on these
materials.
It will be an experimental
challenge to confirm   the structures predicted here.
A difference  NMR spectrum of two identical samples; 
one treated with pressure and the other not,
may give some evidence for the new phases.
Similar difference measurements by ESR and Raman could be 
equally valuable.

We thank D. A. Neumann, R. L. Cappelletti,
and J. E. Fischer for many fruitful discussions  
and critical reading of the  manuscript.
This work  is partially supported by the National Science
Foundation under Grant No. INT97-31014 and
T\"{U}B\.{I}TAK under Grant No. TBAG-1668(197 T 116).

\newpage

{\bf Table 1 }
{\small 
Various structural  parameters and the total energies of the optimized
structures of  (n,0) vdW lattices   and one dimensional interlinked 
(1DI) nanoropes  as shown in Fig.~1. The band gaps of all the structures
listed here are found to be zero, indicating mettallic behavior.}

\bigskip
\noindent
\begin{small}
\begin{tabular}{|c||cc|cc|cc|cc|} \hline\hline
Properties & (5,0)  & (5,0) (1DI) & (6,0)  & (6,0) (1DI)
& (7,0)  & (7,0) (1DI) & (9,0)  & (9,0) (1DI) \\  \hline  \hline
Formula & C$_{20}$  & C$_{20}$  & C$_{24}$ & C$_{24}$ & C$_{28}$ 
& C$_{28}$& C$_{36}$ &C$_{36}$ \\
 Space Group          &Cmcm  &Cmcm  &P6/mmm&Cmcm &Cmcm 
 &Cmcm &P63/mcm&Cmcm \\
 a = b (\AA)              &7.408 &7.079 &8.364 &7.762 
 &9.250&8.432  &10.389 &9.532  \\
 c (\AA)              &4.208 &4.190 &4.212 &4.223 
 &4.218&4.205  &4.219 &4.209  \\
$ \gamma$             &72.46 &125.39&120.00&121.96
&102.35&119.15&120.00 &111.97  \\
Density (gr/cm$^{3}$) &1.8119&2.331 &1.8713&2.2176
&1.5849&2.1415&1.8213 &2.0245  \\ \hline
Energy/C (eV)   & -155.694&-155.802&-155.843&-155.868
&-155.946&-155.969&-156.049 &-156.042  \\
\hline\hline
\end{tabular}
\end{small}

\newpage
\begin{figure}
\caption{
Optimized structures of the  vdW
(6,0) (a),  (7,0) (b),    and one dimensional
interlinked (7,0) (c) nanotube lattices.
The interlinked  structure shown in (c) has lower energy
than vdW packed (7,0) nanotubes shown in (b). }
\end{figure}

\begin{figure}
\caption{
(a) A view along c-axis of  the 1D interlinked  (n,0) nanotube lattice.
The shaded rectangular region is  the orthorhombic unit cell.
Local structure of  carbon atoms involved in the
intertube bonding (dotted lines)  between two (6,0) (b) and (7,0) (c) nanotubes. 
The sp$^{3}$ hybridization occurs between two hexagonal
faces for (6,0) nanotubes  and between a hexagonal face and
a C-C bond for  (7,0) nanotubes.
}
\end{figure}

\begin{figure}
\caption{
Planer lattice constant variation of the total energy 
of (7,0) nanotube ropes in three different phases.
Inset shows the view of the structures along
c-axis. 
The zero of energy was taken 
to be the energy of  vdW packing of the nanotubes.
}
\end{figure}

\begin{figure}
\caption{
Various high density phases of  carbon nanotubes.
(a) Two dimensional interlinked  (2DI) structure   of  
(5,0) nanotubes, consisting of   rectangularly  distorted 
nanotubes  interlinked on a 2D network. 
(b) A hexagonal network of (9,0) nanotubes, in which
(9,0) nanotubes are interlinked along a,b, and [110]
directions.  (c) A very  dense 
structure of (7,0) nanotubes obtained under
30 GPa pressure.  (d) The  optimized
structure of (6,6) nanotubes under P=53 GPa. 
Nanotubes  are distorted
in such a way that the local nearest neighbor structure 
is somewhat similar to  graphite sheets.
$d_{C-C}$ indicates the  smallest distance
between two carbon atoms of nearest neighbor nanotubes.
}
\end{figure}

\newpage
\centerline{\psfig{figure=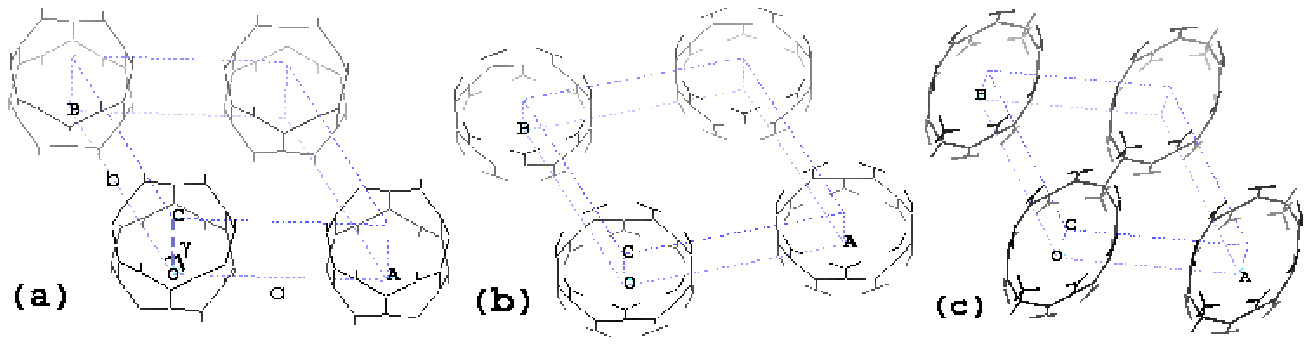,width=170mm}}

\vfill
{\bf Figure 1 } {\bf Yildirim} {\bf\it et al} 

\newpage
\centerline{\psfig{figure=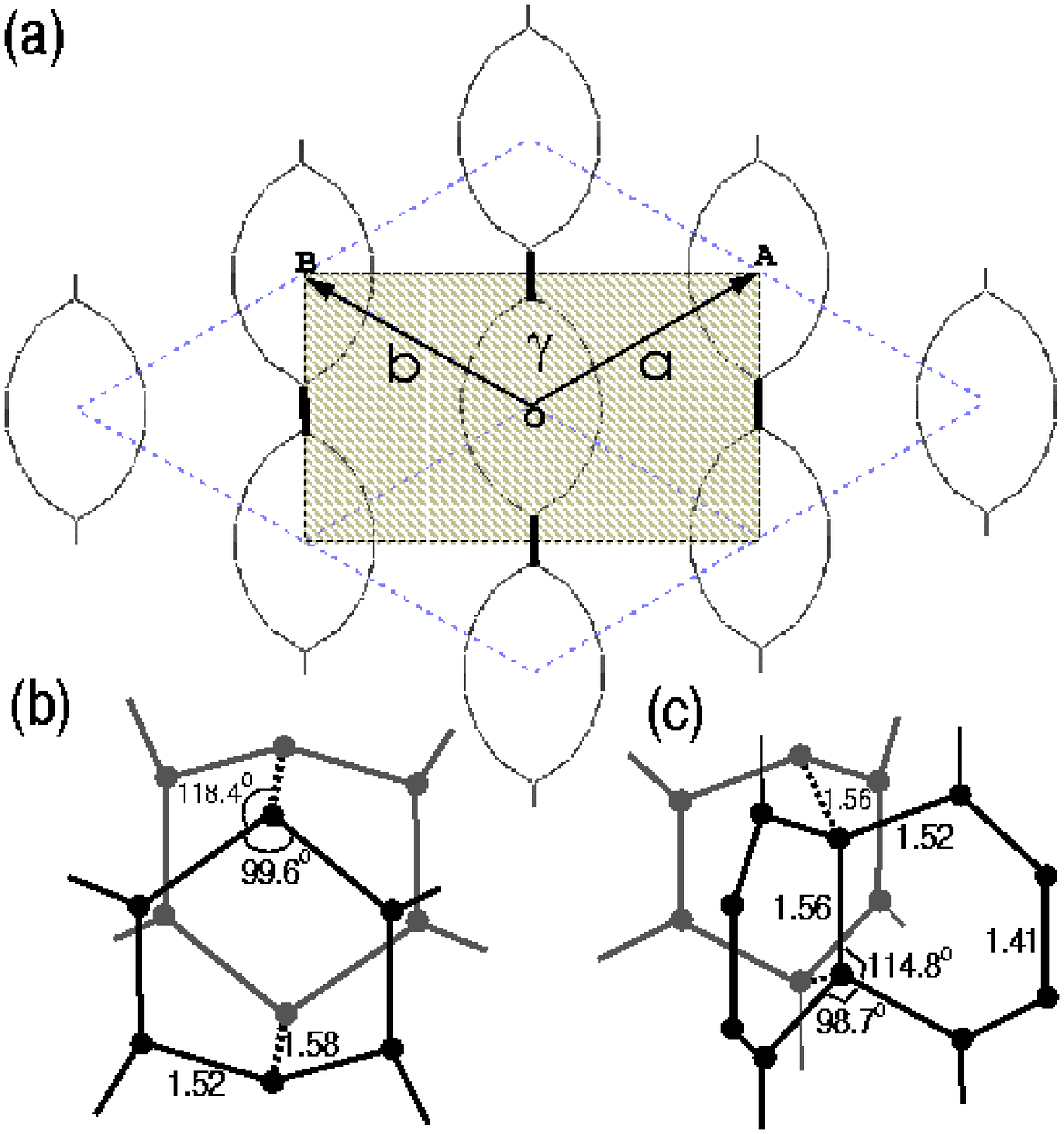,width=150mm}}

\vfill
{\bf Figure 2 } {\bf Yildirim} {\bf\it et al} 

\newpage
\centerline{\psfig{figure=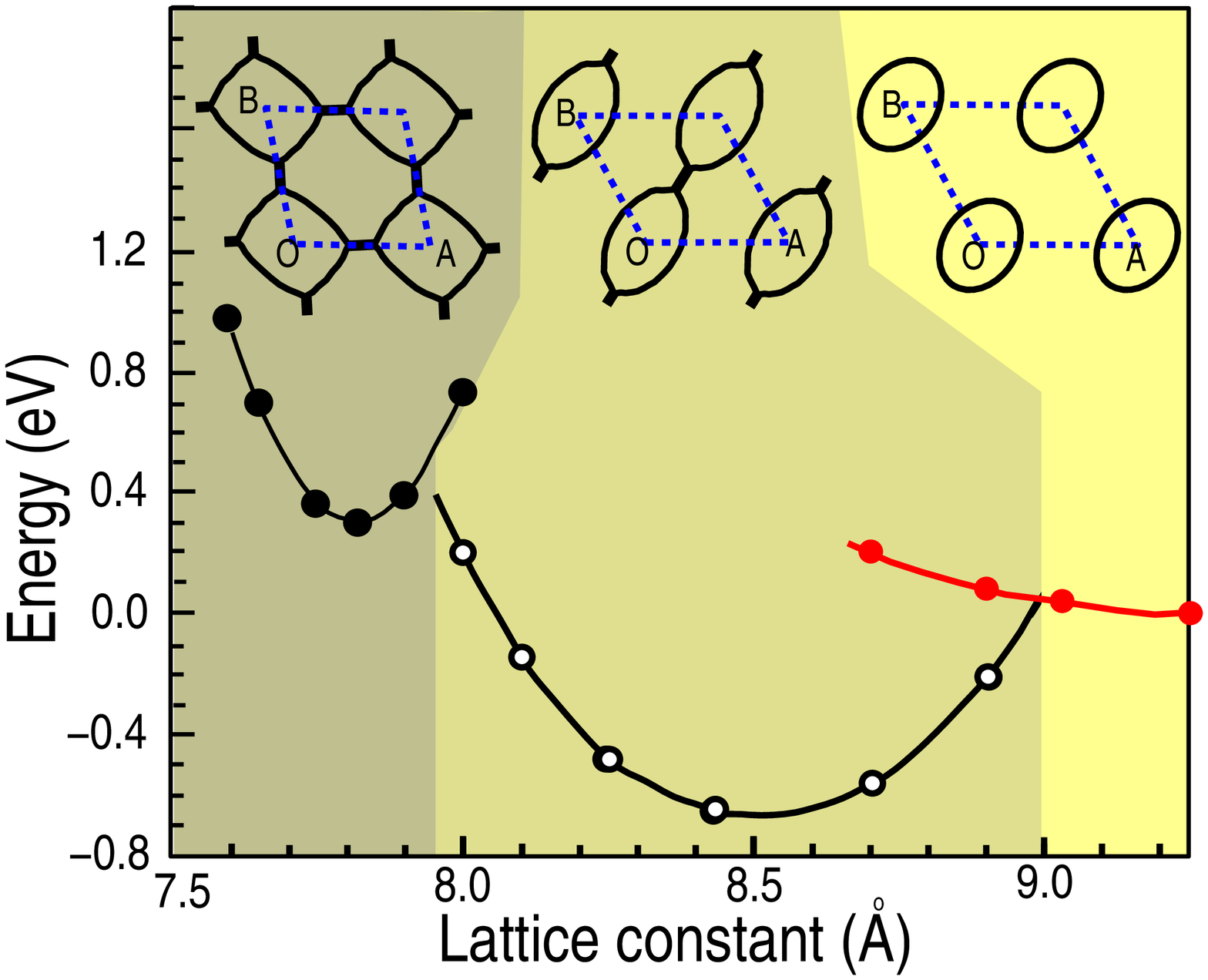,width=150mm}}

\vfill
{\bf Figure 3 } {\bf Yildirim} {\bf\it et al} 

\newpage
\centerline{\psfig{figure=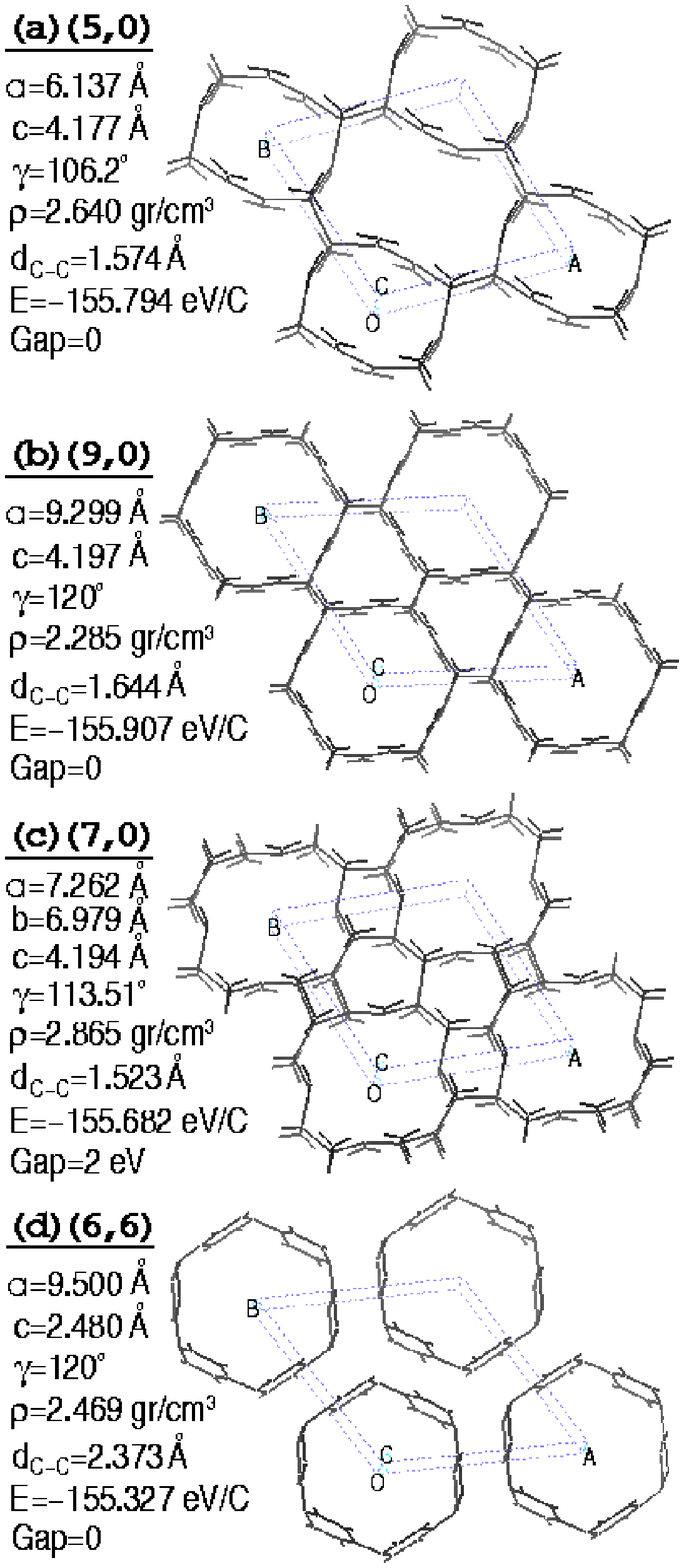,width=85mm}}

\vfill
{\bf Figure 4 } {\bf Yildirim} {\bf\it et al}


\begin{references}
\bibitem{barrier}
We note that the transition path should be calculated
in order to obtain the actual barriers between the 
different phases, which is not done in this paper.
\end{references}
\end{document}